\documentstyle[sig,ricop,12pt]{article}

\begin{document}
\title{Combinatorial tools for Regge Calculus}

\author{E.Fabri}
\address{Dipartimento di Fisica, Universit\`a di Pisa (Italy)}
\author{R.Giannitrapani}
\address{Dipartimento di Fisica, Universit\`a di Trento (Italy)\\
        I.N.F.N gruppo collegato di Trento.}

\maketitle 

\abstract{In this short note we briefly review some recent
mathematical results relevant to the classical Regge Calculus
evolution problem.} 

\section{Introduction} 

Although Regge Calculus \cite{regge}, that is a ``discretized
formulation'' of General Relativity, is now more than 35 years old,
there is an incredible lack of classical solutions obtained with it,
except for highly symmetric ones (Schwarzschild or cosmological
model). This fact is quite unsatisfactory because one of the
motivations of Regge
Calculus was the need for a way to obtain solutions of
General Relativity, even approximate ones, without any particular
symmetry. The origin of this situation is the lack of a general
evolution scheme powerful enough to deal with generic (not symmetric)
spacetime; in this work we review recent results that seem to point
toward a solution of the problem \cite{barr,tuck,sorkin,tesi,tutti}.
Since this is a short communication, we will omit all 
demonstrations and will skip a lot of technicalities; we hope to
present an expanded version of this work in a future paper.

\section{Mathematical tools} 

Let $\cM$ be a ``Regge spacetime'', i.e. a {\em regular simplicial
  complex} \footnote{Here regular means that every $(n-1)$-dimensional
  simplex is a face of a $n$-dimensional simplex.} of dimension $n$
  with a Lorentzian ``metric'', i.e. a particular assignment of a
  length to every edge of $\cM$ (see \cite{sorkin,tesi} for details);
  for the sake of simplicity we consider only compact spacetime. A
  {\em spatial section} $\cN$ is a closed regular $(n-1)$-dimensional
  simplicial subcomplex of $\cM$ with all the edges of ``spatial''
  type; another spatial section $\cN'$ is said to be evolved from
  $\cN$ if every vertex of $\cN'$ is contained in the envelope of
  future cones of the points of $\cN$. We say that such evolution is
  performed by a {\em move} if the subcomplex $\cH \subset \cN$
  composed of the $(n-1)$-simplices that belongs to $\cN$ but not to
  $\cN'$ has the topology of an $(n-1)$-dimensional ball.  If one
  defines analogously $\cH' \subset \cN'$, it is simple to see that
  every move is characterized by a compact $n$-dimensional simplicial
  complex $\cK$ whose boundary is $\cH \bigcup \cH'$; one can think of
  the
  move as the ``gluing'' of $\cK$ on $\cN$.

A $p$-simplex in $\cK$ (with $p < n$) is said to be {\em new} if it
does not belong to $\cH$ and it is said to be {\em internal} if it
does not belong to $\cH'$. It is useful, as we will see later, to
introduce the two quantities $N_p^\cK$ and $I_p^\cK$ that are,
respectively, the number of new $p$-simplices and of internal
$p$-simplices introduced by the move $\cK$.

\smallskip

Sorkin \cite{sorkin} (without using this formalism) has introduced a
very useful move, recently rediscovered and generalized by Tuckey
\cite{tuck}: if $\a$ is a vertex of $\cN$ one introduces a new vertex
$\a'$ in the future cone of $\a$ and than connects it with all the
vertices of $\cN$ that are connected to $\a$ itself. It is easy to see
that this is actually a move and that $\forall p\colon N_p^\cS =
I_p^\cS $, where we have called $\cS$ the Sorkin
move. We will see that such a move plays a fundamental role in the
evolution problem for Regge Calculus.

\smallskip

If $\cK$ is just an $n$-simplex then the move is said to be {\em
elementary} or {\em simple}; obviously there are exactly $n$ different
simple moves that corresponds to the possible ways to glue an
$n$-simplex to an $(n-1)$-dimensional section. The simple moves can be
labeled with the numbers of $(n-1)$-faces of $\cK$ that are in $\cN$;
we use the term $f$-move to indicate a simple move with $f$ such
faces. Indicating with $f$ an $f$-move, it is possible to prove
the following formulas \footnote{We define $\left( \begin{array}{cc}
a\\ b \end{array} \right) = 0$ for $b>a$.}:

$$ N_p^f = \left( \begin{array}{cc} n+1-f \\ n-p \end{array} \right)$$

$$ I_p^f = \left( \begin{array}{cc} f \\ n-p \end{array} \right).$$

It is a well established theorem \cite{barr} that every move $\cK$ can
be decomposed into a sequence of elementary ones; we call
$\Sigma_f(\cK)$ the number of $f$-moves contained in the decomposition
of $\cK$.  Using the preceding formulas we can obtain very easily the
decomposition of a Sorkin move into simple ones; since

$$N_0^\cS = I_0^\cS = 1 $$
we have

$$\Sigma_1(\cS) = \Sigma_n(\cS) = 1 $$
In the same way one obtains

$$\Sigma_m(\cS) = \Sigma_{n+1-m}(\cS).$$

\section{The Regge evolutive problem} 

\subsection{The general case}

The evolution problem in classical Regge Calculus can be formulated in
the following way \cite{barr,tuck,sorkin,tesi,tutti}. 

The starting point is an $n$-dimensional simplicial complex $\cI$ that
has to provide the initial conditions for the problem; all the lengths
of the edges of $\cI$ are known and fixed. The boundary of $\cI$ must
be composed by at least one spatial section $\cN_0$ (and maybe by
another spatial section in the past of $\cN_0$); the problem is to
give an algorithm to evolve $\cN_0$ in time using Regge's
equations. First of all perform a Sorkin move on a vertex $\a \in
\cN_0$, obtaining in this way a new section $\cN_0'$ with the same
connectivity as $\cN_0$. Such a move introduces (see preceding
section) the same number $x_\a$ (that is the number of vertices in
$\cN$ connected to $\a$ plus one) of new edges and of internal edges;
it is easy to see that new edges introduce new unknowns (their
lengths) whereas internal edges introduce new Regge equations (we remind
the reader that Regge equations are associated to the edges of the
lattice).

Not all of these equations are independent because of a sort of {\em
diffeomorphisms invariance} for Regge Calculus
\cite{tesi,galassi}. There are actually $n$ degrees of freedom in the
choice of $n$ lengths of the new edges; the remaining $x-n$ lengths
are fixed by solving an $(x_\a-n)$ dimensional system of equations
associated to $(x_\a-n)$ internal edges \footnote{One can show that
$(x_\a-n)$ is always a positive number if $\cM$ is a regular complex.}
(we call it the {\em Sorkin system}).

After the evolution of $\a$ one can iterate the procedure and evolve,
in the same way, another vertex of $\cN_0$ until all the vertices
of $\cN_0$ have been evolved and one has obtained a new spatial
section $\cN_1$.

This evolving procedure is quite flexible and is useful in a general,
not symmetric case \footnote{It permits also a parallel evolution of
$\cN_0$, but this topic is outside the scopes of this note.}.

Could the decomposition of a Sorkin move into elementary
ones, discussed in the preceding section, further simplify such a
scheme? The answer is yes and we will show briefly this fact in the
next sections.

\subsection{$(2+1)$-gravity}
If $n=3$ one obtains immediately the decomposition of a Sorkin move into
elementary ones: 

$$ \Sigma_1(\cS) = \Sigma_3(\cS) = 1 $$
$$ \Sigma_2(\cS) = x_\a-3.$$
The first move is a $1$-move that introduces $3$ new edges without any
equation, so that there is freedom in the choice of the $3$ new lengths
(lapse and shift freedom of choice); than one performs $(x_a-3)$
$2$-moves, each one introducing one unknown and one equation (since
$N_1^2 = I_1^2 = 1$). Finally one performs a $3$-move that introduces
$3$ equations but no unknowns and so they have to be automatically
satisfied. This shows that the decomposition into elementary moves
permits the complete decoupling of the system of equations related to
a Sorkin move; this fact obviously introduce a big simplification in
numerical uses of Regge Calculus.

\subsection{Some issues for the $(3+1)$-dimensional case}

In $4$ dimensions the decomposition of $\cS$ into elementary moves is a
little more complicated; 

$$ \Sigma_1(\cS) = \Sigma_4(\cS) = 1 $$
$$ \Sigma_2(\cS) = \Sigma_3(\cS) = x_a-4.$$ 
The first and last moves are as before, but now the $2$-moves
introduce each one unknown but no equation ($N_1^2=1$, $I_1^2 = 0$),
while the $3$-moves act in the opposite way ($N_1^3=0$, $I_1^3 = 1$).
In general it is not possible to decouple the Sorkin system of
equations; but it may happen, for particular lattices, that the
Sorkin move is decomposable in a way such that after a $2$-move there
is always a $3$-move. In this particular case one obtain again a
complete decoupling of the Sorkin system of equations, with obvious
numerical advantages. One could imagine also an intermediate situation
with only a partial decoupling of such a system.

We have actually built some explicit examples of lattices with
complete and partial decoupling, but there is not enough space here to
discuss them; we are still investigating the subject.

\section{Conclusion}
It is not possible to discuss here, from a physical point of view, the
results obtained, but it is clear that simplifications in the
evolution problem of the kind we have presented in this short note
could be very useful in an extensive use of Regge Calculus as a
numerical tool of investigation; we hope that this and other
discussions on the subject (as the ones cited in this work) can open a
new ``golden age'' for classical Regge Calculus.

\section*{Acknowledgments}
One of us (R.G.) is grateful to R.M.Williams and R.D.Sorkin for useful 
conversations during a Workshop in Vietri (Italy) in the summer of 1995; 
he thanks also P.A.Tuckey, J.W.Barrett, M.Galassi and A.Barbieri 
for invaluable suggestions.

\end{document}